\documentstyle[preprint,aps]{revtex}
\title{Virtual Pion Scattering}
\author{Marios A. Kagarlis}
\address{Niels Bohr Institute, Blegdamsvej 17, DK 2100 Copenhagen, Denmark}
\author{Vladimir F. Dmitriev}
\address{Budker Institute of Nuclear Physics, Novosibirsk-90, 630090
Russia}

\begin{document}

\maketitle

\begin{abstract}
We propose a theory which exploits the charge-exchange reactions
($^3$He,$^3$H$\pi^+$) and (p,n$\pi^+$) as effective sources of virtual
pions.
We consider processes in which the creation of virtual pions is followed by
conventional coupled-channel pion scattering to discrete nuclear states.
This
picture allows us to incorporate successful theories of pion scattering and
utilize virtual pions as probes of the nuclear matter. For coherent pion
production we clearly demonstrate that the shift of the coherent peak
position in the excitation function of $^3$He-A
relative to $^3$He-N scattering is determined entirely by the pion
nucleus rescattering.

\end{abstract}


The effect that the presence of nuclear matter has on reactions of nucleons
and mesons remains one of the most formidable open questions in nuclear
physics. For many years pion beams have provided a unique probe for the
study
of nuclei with considerable success. Lately, the possibility has been
raised that virtual pions may also be useful as a complementary
probe\cite{me}.
Coherent pion production in nuclei has often been likened to elastic pion
scattering\cite{eot1} and may be utilized as an effective virtual pion
source.
It comprises of a charge-exchange reaction and the simultaneous creation of
a
pion, leaving the target nucleus in its ground or a low-lying excited
state.
Although it is a small channel relative to inclusive pion production, it
has
recently attracted a lot of attention\cite{th93,po93,vd93,pfdc93,jd91}.
Here
we present a theory of coherent pion production which we believe is
especially suitable
for the study of the medium effects, and can shed light on the nature
of the interaction between virtual pions and nuclear matter in this
reaction.

One persistent feature established from measurements of cross sections in
the
charge-exchange reactions ($^3$He,$^3$H$\pi^+$) and (p,n$\pi^+$) is a
downward
shift in the transfer energy spectra of the $^3$He-nucleus and p-nucleus
relative to the $^3$He-nucleon and p-nucleon
reactions\cite{th93,vga84,dc86,cgc76,th92}. Extensive
theoretical work has covered many aspects of these interactions both in the
$\Delta$-hole model\cite{po93,vd93,jd91}, with emphasis on the nuclear
response function, as well as in the picture of \cite{eot1,pfdc93,eo89}
with emphasis on the elementary reaction. These models reproduce reasonably
well the measured cross sections but do not reach consensus on the nature
of the medium effects. In particular, none of the previous models provides
an
unambiguous interpretation of the observed energy shift, although the
picture
employed by \cite{eot1} is in principle similar to ours. In our model the
energy shift acquires a clear physical interpretation, and the connection
between coherent pion production (to the g.s. and low-lying excited states
of
the target nucleus) and pion scattering (elastic and inelastic) is
established
quantitatively. Furthermore, our theory sets limits on the sensitivity that
future coherent pion production experiments \cite{soc} must attain, and
the sort of information one would hope to extract from them.

We employ a framework which views coherent pion production as a two-step
process: creation of a virtual pion, followed by scattering of the pion
with
nucleons in the target nucleus. The former is treated formally as a source
term (effective virtual pion beam) while the latter is treated as pion
scattering via a pion-nucleus optical potential. This is the main
        difference between our approach and the previous calculations
\cite{po93,pfdc93} where the main focus was on the
$\Delta$ - hole dynamics determined by a meson exchange interaction
with the exchange by $\pi$- and $\rho$-mesons. In our picture the
$\Delta$ does not have explicit dynamical degrees of freedom, it is
hidden in a resonant parametrization of pion-nucleus optical potential.
Apart from resonant contribution the pion-nucleus optical potential has a
contribution from s-wave pion nucleon scattering amplitude which is
important at pion c.m.  kinetic energy below 50 MeV.

 We consider coherent pion production with
$^3$He and p as probes. This formulation is concisely stated in the
inhomogeneous coupled-channel Klein-Gordon equations, which we solve in
coordinate space:
\begin{eqnarray}
&&\left(\frac{1}{r^2}\frac{d}{dr}r^2\frac{d}{dr} -
\frac{l_{\beta}(l_{\beta} +
1)}{r^2} - U^{l_{\beta}I_{\beta}}_{l_{\beta}I_{\beta}J}(r)
+ k^2_{\beta} \right)R^{l'I'J'}_{l_\beta I_\beta J}(r)=\nonumber\\
&&\sum_{\alpha} U^{l_\alpha I_\alpha}_{l_\beta I_\beta J}(r)R^{l'I'J'}_{
l_\alpha I_\alpha J}(r) + \rho^{l'I'J'}_{l_\beta I_\beta J}(r)\ .
\label{ikg}
\end{eqnarray}
Note that $l, I$ and $J$ are pion, nucleus and coupled-channel angular
momenta
and the indices $\alpha, \beta$ denote initial and final states involving
real
pions, whereas prime denotes the entrance channel with a virtual pion.
The explicit separation of the source term $\rho^{l'I'J'}_{l_\beta I_\beta
J}(r)$ from the remaining (homogeneous) Klein-Gordon equation for
pion-nucleus
scattering
enables us to disentangle the pion production from the pion interaction
with
the medium, and better understand how each component participates in the
reaction. Similar approach has been successfully used first in nuclear
reaction theory to describe the transfer reactions \cite{GlAs69}, and
for the $\Delta$ - hole system in calculations of the $\Delta$- hole
response function \cite{uho90}, and a pionic decay of the correlated
$\Delta$-hole state leading to coherent pions \cite{po93}.

 While the details will be provided elsewhere\cite{kd95}, let
us briefly discuss Eq.~(\ref{ikg}): The pion-nucleus optical potential
\begin{equation} U^{l_\alpha I_\alpha}_{l_\beta I_\beta J}(r) =
U_0\delta_{\alpha\beta} + <l_\beta I_\beta J|V_{\pi A}|l_\alpha
I_\alpha J> \label{op} \end{equation} has been applied successfully for
elastic, inelastic and charge exchange pion-nucleus scattering (with
the homogeneous Klein-Gordon equation). The diagonal component includes
medium corrections which have been shown to be well-described in terms
of pion absorption, the Lorentz-Lorenz effect, and Pauli correlations
for low pion energies (see \cite{mk1,mk2} and references therein),
while near resonance it includes phenomenological corrections from fits
to elastic, inelastic and double-charge exchange pion-nucleus data (see
\cite{mk3} and references therein). The pion-nucleus transition operator
$V_{\pi A}$ is derived microscopically from the pion-nucleon t-matrix,
summed
over all the valence nucleons and weighed by one-body transition densities
from the shell model. Its matrix elements between pion-nucleus
coupled-channel
states, as in Eq.~(\ref{op}), are the off-diagonal transition matrix
elements.

The advantages of this formalism -- as explained in Refs.~\cite{mk1,mk2} --
include taking into account the internal and external distortions of the
pion
waves, avoiding the closure approximation by explicitly including the
excitation energy of all the nuclear states, and using nuclear structure
input
from shell-model calculations. All these features were shown to be
important
in pion charge-exchange scattering. In particular, the external pion
distortions were necessary for the physical behaviour of the model while
the
other improvements accounted for corrections.

For the case at hand, of equal importance is the evaluation of the source
term
$\rho^{l'I'J'}_{l_\beta I_\beta J}(r)$. We derive a microscopic
coordinate-space operator for coherent pion production, which includes
corrections due to the Fermi motion of the nucleons in the target nucleus,
starting from the t-matrix\cite{kd95}
\begin{eqnarray}
&&t_{NN\rightarrow NN\pi} = 4\pi {{\hbar c}\over{m_\pi^3}}f_{\pi NN}
f_{\pi N\Delta}^2 t'_{N\Delta}\left({ {{\Lambda'}_\pi^2 - m_\pi^2}\over
{{\Lambda'}_\pi^2 - t} }\right)^2\nonumber\\
&& \frac{\left[ (\hat{\sigma}\cdot\hat{q})\,
(\hat{S}^{\dag}\cdot\hat{q}) +
(\hat{\sigma}\times\hat{q})\cdot(\hat{S}^{\dag}
\times\hat{q})
\right]}{\omega - E_{res} + \imath \frac{\Gamma (\omega)}{2}}
\,(\hat{S}\cdot\hat{k})\,\hat{T}^{\dag}\,(\hat{T}\cdot\hat{\tau})\ ,
\label{tmatrix}
\end{eqnarray}
where $t'_{N\Delta}$ = 0.6 \cite{po93} and $\Lambda'_\pi$ = 650~MeV
\cite{po93,vd86}. We choose the particular model for
pion production amplitude via $\Delta$ in order to be able to compare the
results with the previous calculations \cite{po93,vd93,pfdc93} .  We
stress that Eq.~(\ref{tmatrix}) must be interpreted as purely
phenomenological, which however roughly reproduce the main observable
features of the elementary reaction $NN\rightarrow N\Delta$, the ratio of
spin-transverse and spin-longitudinal cross-sections close to 2:1, and
weak energy dependence at intermediate energies.

With the coherent pion production operator as described we obtain the
source term $\rho^{l'I'J'}_{l_\beta I_\beta J}(r)$ using a distortion
factor
for the projectile (in the case of $^3$He) in the eikonal approximation
following the treatment of Refs.~\cite{vd93,vd89}.

We have applied this model for calculations of coherent pion production
(program MEGAPI\cite{kd95}) with a $^3$He probe on a $^{12}$C nucleus to
the
ground state (g.s.) as well as, for the first time, to the lowest-lying
2$^+$ and 1$^+$ states (Fig~1). As the resolution of future experiments
will be sufficient to distinguish between the g.s. and low-lying
excited states, these results are instructive\cite{soc}. All our
calculations were performed with harmonic oscillator wavefunctions for
the nucleons, except for one (dashed) in Fig~1a with Hartree-Fock
wavefunctions\cite{hf}, in order to test the sensitivity of our model
to the nuclear density input. The one-body transition densities which
we used both in the source as well as in the pion scattering terms of
Eq.~(\ref{ikg}) were obtained from the shell-model code OXBASH\cite{ox}
with
the Cohen-Kurath interaction. We also show calculations with p as a probe
on
$^{12}$C and $^{40}$Ca to the g.s. (Fig~2). Canonical calculations to the
g.s.
include the source term (production) and the diagonal piece of the
pion-nucleus optical potential Eq.~(\ref{op}) (pion rescattering) with all
its
higher-order medium corrections as described earlier, plus pion
rescattering via intermediate excited states via the off-diagonal piece
of potential.  For the excited states, canonical calculations include
direct contributions from the source to the g.s.  as well as the
excited states, followed by rescattering via the diagonal pion-nucleus
optical potential and medium corrections, and scattering from the g.s.
to excited states via the off-diagonal component of Eq.~(\ref{op}).

The calculations for $^{12}$C are consistent with the available data (see
e.g.
Ref.~\cite{th93}). Our findings concerning the interaction of the virtual
pion
with the nucleus are contained compactly in Fig~1c: The dashed curve is a
calculation where the optical potential of Eq.~(\ref{op}) has been omitted.
It is equivalent to ``removing" the nuclear medium, except for a residual
presence as a density of finite volume of the target nucleus in the source
term. In this case the cross-section to the g.s. is proportional to the
target formfactor squared. The target formfactor is a Fourie
component $\rho ({\bf q - k})$ of the density present in the source
term. Here ${\bf q}$ and ${\bf k}$ are the incoming momentum transfer
from the projectile and the outgoing pion momentum. The formfactor
defines the pion angular distribution width at high pion momentum.
At low pion momentum the formfactor suppresses the pion yield due to
momentum mismatch which is the largest at low pion energy.
 The dot-dashed curve includes pion rescattering via a
lowest-order pion-nucleus optical potential without the higher order
medium corrections discussed earlier. This results in a dramatic
reduction of the total cross section near the $\Delta$ resonance --
corresponding to T$_\pi\sim$160~MeV, $\omega\sim$300~MeV -- where
p-wave pion scattering dominates, and a small enhancement at lower pion
energies, in the range of s-wave dominance. The reduction comes from
a resonant pion absorption with a subsequent incoherent $\Delta$
decay. The incoherent decay width of the $\Delta$ is large
compared to the decay width into a single coherent channel and this reduces
the yield of coherent pions near the $\Delta$ peak. Thus,
 the downward shift in this approach is determined mainly by the
resonance absorption mechanism. This interpretation of the peak
position differs from previous calculations \cite{po93,pfdc93} where
the peak position was determined by the $\Delta$-nucleus dynamics.
  Finally, the solid curve is
a canonical calculation as described earlier, with a pion-nucleus
optical potential and higher order medium corrections as determined
from pion-nucleus scattering.  The difference between the dot-dashed
and the solid curves reflects the details required of future coherent
pion production experiments in order to probe the higher order
corrections of the pion-nucleus optical potential, and potential
differences between real and virtual pionic probes.

In summary, we have presented a theory of coherent pion production which
utilizes this reaction as an effective virtual pion source and probes the
nature of the interaction between virtual pions and the nucleus. We have
established the close connection between coherent pion production and pion
scattering by showing the considerable change in the energy spectra of
$^3$He-A and p-A relative to $^3$He-N and p-N as a result of pion
nucleus rescattering.
The explicit separation of pion production
 from rescattering in our theory provides a handle for the study of
 the higher-order medium effects in the production amplitude, which
illustrate the range of possible differences between real and virtual
pionic probes.

This work was supported in part (V.F.D.) by International Science
Foundation, grant NQE000. Marios A. Kagarlis acknowledges funding from
the European Union program ``Human Capital and Mobility", and thanks
Eulogio Oset for illuminating discussions and criticism.  Both the
authors are indebted to Carl Gaarde for his invaluable support at all
the stages of this project.

\figure{The ($^3$He,$^3$H$\pi^+$) calculations, for T$_{^3He}$ = 2 GeV
and $\theta_{^3H}$ = 0 deg: a) Angular distribution for charge-exchange on
a
$^{12}$C target left in the g.s. and transfer $\omega$ = 260 MeV, with a
Woods-Saxon (solid) and Hartree-Fock (dashed) density.
b) As before but for the lowest-lying 2$^+$ (solid) and 1$^+$ (dashed)
states in $^{12}$C. c) The $\omega$-transfer dependence of the total
cross section for the g.s. transition in $^{12}C$ without rescattering
of the coherent pion (dashed), with rescattering via a lowest-order
pion-nucleus optical potential (dot dashed), and with rescattering
via the full pion-nucleus optical potential including higher-order
medium effects (solid). d) Full calculations of the $\omega$-transfer
dependence for the 2$^+$ (solid) and 1$^+$ (dashed) states in $^{12}$C.}
The irregularity of the spectra at 260 - 280 MeV comes from use of
different
parametrization of pion-nucleus optical potential at low and high pion
energy.

\figure{The (p,n$\pi^+$) calculations, for T$_p$ = 800 MeV and
$\theta_n$ = 0 deg, for charge-exchange transitions to the ground state:
a) Angular distribution for transfer $\omega$ = 260 MeV on a $^{12}$C
target. b) As before but for $^{40}$Ca. c) $\omega$-transfer dependence
for charge-exchange without rescattering (dashed), and with rescattering
via the full pion-nucleus optical potential (solid) on a $^{12}$C target.
d) As in (c) but for $^{40}$Ca.}

\end{document}